\documentclass[11pt]{article}
\usepackage{amssymb}
\usepackage{amsmath}
\usepackage{amsthm}

\usepackage{graphicx,color}

\newcommand{\ket}[1]{\left | #1 \right \rangle}
\newcommand{\bra}[1]{\left \langle #1 \right |}

\def\ra{\rangle}

\def\openone{\leavevmode\hbox{\small1\kern-3.8pt\normalsize1}}

\def\cl{{\cal L}}

\def\cn{{\cal N}}

\def\cg{{\cal G}}
\def\ca{{\cal A}}

\def\cc{{\cal C}}

\def\cm{{\cal M}}
\def\cs{{\cal S}}
\def\cu{{\cal U}}

\def\RR{\mathbb{R}}

\def\CC{\mathbb{C}}

\newtheorem{theorem}{Theorem}
\newtheorem{lemma}{Lemma}

\newtheorem{corollary}{Corollary}

\theoremstyle{definition}

\newcommand{\proj}[1]{\ket{#1}\!\bra{#1}}

\begin{document}
\title{\LARGE\bf
 Jordan-Wigner formalism for\\ arbitrary 2-input 2-output matchgates\\ and their classical simulation}
\author{Richard Jozsa$^1$,
  Akimasa Miyake$^2$ and Sergii Strelchuk$^1$\\[3mm]
  \small\it
  \small\it $^1$DAMTP, Centre for Mathematical Sciences, University of Cambridge,\\ \small\it Wilberforce Road, Cambridge CB3 0WA, U.K.\\[1mm]
  \small\it $^2$Center for Quantum Information and Control, Department of Physics and Astronomy, \\ \small\it
University of New Mexico, 1919 Lomas Blvd NE,  Albuquerque NM 87131, USA.}

\date{}

\maketitle

\begin{abstract}
In Valiant's matchgate theory, 2-input 2-output matchgates are $4\times 4$ matrices that satisfy ten so-called matchgate identities. We prove that the set of all such matchgates (including non-unitary and non-invertible ones) coincides with the topological closure of the set of all matrices obtained as exponentials of linear combinations of the 2-qubit Jordan-Wigner (JW) operators and their quadratic products, extending a previous result of Knill. In Valiant's theory, outputs of matchgate circuits can be classically computed in poly-time. Via the JW formalism, Terhal \& DiVincenzo and Knill established a relation of a unitary class of these circuits to the efficient simulation of non-interacting fermions. We describe how the JW formalism may be used to give an efficient  simulation for all cases in Valiant's simulation theorem, which in particular includes the case of non-interacting fermions generalised to allow arbitrary 1-qubit gates on the first line at any stage in the circuit. Finally we give an exposition of how these simulation results can be alternatively understood from some basic Lie algebra theory, in terms of a formalism introduced by Somma et al.

\end{abstract}

\section{Introduction}\label{intro}
The theory of matchgate computations was introduced by Valiant in \cite{val2} and used to provide a striking new class of efficient (i.e. poly-time) classical algorithms for a variety of computational tasks \cite{val1}. General matchgates can have $k$ inputs and $l$ outputs, being then represented by matrices of size $2^k\times 2^l$. Here we will be concerned only with 2-input 2-output matchgates and hereafter the term `matchgate' will always mean `2-input 2-output matchgate'. 

In some cases, matchgates can be unitary and Valiant also identified a corresponding novel class of classically efficiently simulatable quantum circuits \cite{val2}. Soon thereafter Terhal \& DiVincenzo \cite{terdiv}  and Knill \cite{knill} showed that there is a connection between a class of Valiant's unitary matchgate circuits and the physics of non-interacting fermions (and see \cite{jm08} for a further exposition).  The evolution of non-interacting fermions can be classically efficiently simulated \cite{brkit,terdiv} by using the formalism of Jordan-Wigner (JW) operators \cite{jw} (that gives a representation of fermionic modes in terms of standard qubits), and this  provided a quantum physical interpretation of part of Valiant's results. In this paper we will further extend and study the relationship between Valiant's matchgate theory and the Jordan-Wigner formalism of quantum physics.

Our results are organised as follows. In Section \ref{sect2} and the Appendix we will prove an equivalence between arbitrary 2-input 2-output matchgates (including non-unitary and non-invertible ones) and the JW formalism for two qubit lines, viz. the set of all such matchgates will be seen to coincide with the topological closure of the set of all matrices obtained as exponentials of linear combinations of the 2-qubit Jordan-Wigner (JW) operators and their quadratic products, extending a previous result of Knill \cite{knill}. Then in Section \ref{sect3} we will see that the classical simulation of all cases of matchgate circuits in Valiant's simulation theorem \cite{val2}, can be carried out using the Jordan-Wigner formalism.  Finally in Section \ref{lie} we will give an exposition of how all these simulation results can also be alternatively understood using some basic Lie algebra theory, as an example of a more abstract formalism introduced by Somma et al. \cite{somma}.

In fermionic physics it is usual to impose conservation of the parity of the number of fermions (the boson-fermion superselection rule). In the JW formalism this corresponds to considering quantum processes whose hamiltonians involve only even degree products of the JW operators, and the case of so-called non-interacting fermions corresponds to allowing only purely quadratic terms. It may be shown \cite{terdiv} \cite{jm08} that in the matchgate formalism, the latter case is equivalent to considering poly-sized circuits of  unitary 2-qubit matchgates of the following form, each acting on two nearest-neighbour (n.n.)  qubit lines:
\begin{equation}\label{gab} G(V,W) = \left(
\begin{array}{cccc} p&0&0&q \\ 0&w&x&0 \\ 0&y&z&0 \\ r&0&0&s
\end{array} \right) \hspace{1cm} V = \left( \begin{array}{cc}
p&q \\ r&s \end{array} \right) \hspace{5mm} W= \left(
\begin{array}{cc} w&x \\ y&z \end{array} \right) .\end{equation}
Here $V$ and $W$ are both in $SU(2)$ or both in $U(2)$ with the
same determinant and ``n.n." is with respect to any fixed chosen linear ordering of the qubit lines. We will refer to gates of this form acting on n.n. qubits as {\em fermionic} matchgates.
However Valiant's matchgate formalism includes further unitary circuits that do not respect the boson-fermion superselection rule; an interesting example is the following: we can have circuits of fermionic matchgates together with arbitrary 1-qubit gates applied on the {\em first} qubit line at any stage within the circuit. We will see (as also outlined in \cite{knill}, and in accordance with our general JW-matchgate equivalence), that they may be represented in terms of hamiltonians that have {\em linear} as well as quadratic JW terms. Furthermore these seemingly more general hamiltonians can in fact be seen as special cases of purely quadratic ones in a slightly enlarged setting.

\section{General matchgates and the JW formalism}\label{sect2}
In this section we will be concerned with just two qubit lines and $4\times 4$ matrices.
We will label rows and columns of $4\times 4$ matrices $B$ by 1, 2, 3, 4 which will correspond respectively to 00, 01, 10, 11 when $B$ is viewed as operating on the space of two qubits.

The Jordan-Wigner operators for $n$ qubit lines are defined in eq. (\ref{jwr}) below and for two qubit lines we have simply
\begin{equation}\label{jws} c_1=XI\hspace{5mm} c_2=YI\hspace{5mm} c_3=ZX\hspace{5mm} c_4=ZY
\end{equation}
where $I$ is the identity and $X,Y,Z$ are the standard Pauli matrices, and we have omitted all tensor product symbols (so e.g. $XI$ is shorthand for $X\otimes I$).

For matchgates, it was shown in \cite{ccl} that a $4\times 4$ matrix $B$ is a 2-input 2-output matchgate if and only if it satisfies the following ten {\em matchgate identities}:
\begin{align*}
M_1 =  B_{11}  B_{44}- B_{14}  B_{41}- B_{22}  B_{33}+ B_{23}  B_{32}=0\\
M_2 =  B_{21}  B_{44}- B_{22}  B_{43}+ B_{23}  B_{42}- B_{24}  B_{41}=0\\
M_3 =  B_{31}  B_{44}- B_{32}  B_{43}+ B_{33}  B_{42}- B_{34}  B_{41}=0\\
M_4 =  B_{13}  B_{44}- B_{14}  B_{43}- B_{23}  B_{34}+ B_{24}  B_{33}=0\\
M_5 =  B_{12}  B_{44}- B_{14}  B_{42}- B_{22}  B_{34}+ B_{24}  B_{32}=0\\
M_6 =  B_{11}  B_{24}- B_{12}  B_{23}+ B_{13}  B_{22}- B_{14}  B_{21}=0\\
M_7 =  B_{11}  B_{42}- B_{12}  B_{41}- B_{21}  B_{32}+ B_{22}  B_{31}=0\\
M_8 =  B_{12}  B_{43}- B_{13}  B_{42}- B_{21}  B_{34}+ B_{24}  B_{31}=0\\
M_9 =  B_{11}  B_{34}- B_{12}  B_{33}+ B_{13}  B_{32}- B_{14}  B_{31}=0\\
M_{10} =  B_{11}  B_{43}- B_{13}  B_{41}- B_{21}  B_{33}+ B_{23}  B_{31}=0
\end{align*}
Let $\cm\cg$ be the set of all 2-input 2-output matchgates, which is thus a closed set in the space of all $4\times 4$ complex matrices.

 We will prove a correspondence between $\cm\cg$ and $4\times 4$ matrices obtained as exponentials $e^A$ of linear combinations of the $c_i$'s and quadratic terms $c_ic_j$'s i.e. $A=\sum \alpha_i c_i+\sum \beta_{ij}c_ic_j$ with $\alpha_i,\beta_{ij}\in \CC$. However to obtain the matchgate identities as written above we will need to reverse the order of the tensor product in the JW operators and use
\begin{equation}\label{wjs} \tilde c_1=IX\hspace{5mm} \tilde c_2=IY\hspace{5mm} \tilde c_3=XZ\hspace{5mm} \tilde c_4=YZ.
\end{equation}
If we were to proceed instead with the JW operators directly (e.g. as was done in \cite{knill} for a setting of five matchgate identities) we would obtain matrices $e^A$ that do not satisfy the identities above, but instead satisfy a set of identities obtained by changing row and column labels via $1,2,3,4 \rightarrow 1,3,2,4$ or $00,01,10,11 \rightarrow 00,10,01,11$ corresponding to reversing the role of the two qubits. For example instead of $M_2=0$ above we would obtain $B_{31}  B_{44}- B_{33}  B_{42}+ B_{32}  B_{43}- B_{34}  B_{41}=0$  (viz. eq. (3) in \cite{knill}).

Thus introduce the eleven linear and quadratic reversed JW operators (up to overall constants
and writing $\tilde{c}_0$ for $\tilde c_i\tilde c_j$ with $i=j$):
\begin{equation}\label{liealg}\begin{array}{cccccc}
\tilde c_0=II & & & & & \\
\tilde c_1=IX & \tilde c_2=IY& \tilde c_3=XZ & \tilde c_4=YZ & & \\
\tilde c_1\tilde c_2=IZ & \tilde c_1\tilde c_3=XY& \tilde c_1\tilde c_4=YY & \tilde c_2 \tilde c_3=XX & \tilde c_2 \tilde c_4=YX & \tilde c_3 \tilde c_4=ZI
\end{array}
\end{equation}
Let $\cl$ be the (complex) linear span of these eleven $4\times 4$ matrices, which is an 11-dimensional Lie algebra (with the standard matrix product commutator as Lie bracket).\\  Let $\cg= \{ e^A:A\in \cl \}$ be the corresponding Lie group. Note that by definition, all elements of $\cg$ are invertible matrices but $\cg$ is not topologically closed; for example
 $\lim_{t\rightarrow \infty} e^{t(IZ-II)} =\lim_{t\rightarrow \infty} e^{t(IZ)}/e^t = I\otimes \proj{0}\notin \cg$, 
or even more simply $\lim_{t\rightarrow \infty} e^{A-tII} = \lim_{t\rightarrow \infty} e^{A}/e^t$ which is the zero matrix.

Let $\cm\cg^*= \{ B\in \cm\cg : \mbox{$B$ is invertible} \}$. Then we have:
\begin{theorem}\label{thm11} 

{\bf (a)} $\cm\cg^*$ is dense in $\cm\cg$;\hspace{3mm} {\bf  (b)} $\cm\cg^* =\cg$.\\ Thus $\cm\cg=\overline{\cg}$ (where the overline denotes topological closure).
\end{theorem}
In the Appendix we give a full proof of this Theorem, utilising and extending some methods introduced in \cite{knill}. The argument contains some further results of possible independent interest e.g. a geometrical interpretation of the matchgate identities, given in Theorem \ref{thm22} in the Appendix, from which it will follow that $\cg$ is a group.

\section{Classical simulation of matchgate circuits}\label{sect3}
We now introduce the full  Jordan-Wigner formalism and describe how it may be used to provide an efficient classical simulation of the matchgate circuits treated in \cite{val2}.

The Jordan-Wigner operators for $n$ qubit lines are the $2n$ Pauli product operators
 (omitting tensor product symbols $\otimes$
throughout):
\begin{equation}\label{jwr} \begin{array}{ccccc}
c_1=X\,I\ldots I & & c_3= Z\,X\,I\ldots I & & c_{2k-1}= Z\ldots
Z\,X\,I\ldots I \hspace{8mm} {\rm etc.}
\\
c_2=Y\,I\ldots I & & c_4= Z\,Y\,I\ldots I & & \,\,c_{2k}\,\,\,\, =
Z\ldots Z\,Y\,I\ldots I \hspace{8mm} {\rm etc.}
\end{array}
\end{equation} Here $k$ ranges from 1 to $n$, and the Pauli $X$ and $Y$ operators are in the $k^{\rm th}$ slot for $c_{2k-1}$ and $c_{2k}$. We say that the operators $c_{2k-1}$ and $c_{2k}$ are associated to the $k^{\rm th}$ qubit line. These $n$-qubit operators are Hermitian and satisfy the Clifford algebra anti-commutation relations:
\begin{equation}\label{cliffcomm}  \{ c_\mu,c_\nu \} \equiv c_\mu c_\nu + c_\nu c_\mu = 2\delta_{\mu\nu} I\hspace{5mm} \mu , \nu = 1, \ldots , 2n. \end{equation}
 We begin by establishing some algebraic properties of Clifford algebras and later we will apply these to the representation provided by the JW operators.

\subsection{Quadratic and linear terms in a Clifford algebra}
Let $c_\mu$, $\mu=1,\ldots ,m$ be any $m$ symbols (now abstract Clifford algebra generators rather than the JW representation above) that satisfy the Clifford algebra anti-commutation relations $ \{ c_\mu,c_\nu \} \equiv c_\mu c_\nu + c_\nu c_\mu = 2\delta_{\mu\nu} I\hspace{5mm} \mu , \nu = 1, \ldots , m$.

Let $\cl_1$ denote the complex linear span of the $c_\mu$'s and let $\cl_2$ be the complex linear span of all purely quadratic terms $c_\mu c_\nu$. Let $\cl_{1\oplus 2}=\cl_1\oplus \cl_2$ be the linear span of all linear and quadratic terms. Note that from the Clifford algebra anti-commutation relations, $\cl_1$, $\cl_2$ and $\cl_{1\oplus 2}$ respectively have dimensions $m$, $\binom{m}{2}+1$ and $\binom{m}{2}+m+1$.

\begin{theorem}\label{thm1} Let $T=e^A$ for $A=\sum_{\mu ,\nu} a_{\mu \nu}\,  c_\mu c_\nu \in \cl_2$ be the exponential of any purely quadratic expression in the $c_\mu$'s. Then $\cl_1$ is preserved under conjugation by $T$. \end{theorem}

\noindent {\bf Proof}\, The terms in $A$ with $\mu = \nu$ contribute only an overall scalar multiple for $T$, giving a trivial conjugation action on $\cl_1$. Thus (recalling also that $c_\mu c_\nu =-c_\nu c_\mu$) we may assume without loss of generality that $a_{\mu \nu}$ is anti-symmetric. For that case,
a simple proof of the Theorem is given in \cite{jm08} (theorem 4.1 there) for the case of $m$ even and $a_{\mu \nu}$ being real and anti-symmetric (guaranteeing there that $A$ is hermitian for hermitian $c_\mu$'s). But it is easy to see that the proof extends without change to the general case of arbitrary anti-symmetric complex $a_{\mu\nu}$'s. For each $c_\mu$ we get $Tc_\mu T^{-1} = \sum_\nu K_{\mu \nu}\,  c_\nu$ with $[ K_{\mu \nu}]=\exp (-4[a_{\mu \nu}])$, where square brackets denote matrices with the given entries.\, $\Box$

\noindent {\bf Remark 1.}\, Note that Theorem \ref{thm1} also implies the preservation of $\cl_2$ since $TMN T^{-1} = (TMT^{-1}) (TNT^{-1})$ for any  $M,N \in \cl_1$ $\Box$.

Below we will be interested in extending the exponent to include linear terms:
\[ A=\sum_{\mu ,\nu} a_{\mu \nu}\, c_\mu c_\nu +\sum_\sigma b_\sigma c_\sigma\,\,\, \in \cl_{1\oplus 2}. \]
But neither $\cl_1$ nor $\cl_2$ is closed under conjugation by these extended $e^A$'s.  However we can view any  such extended exponent as a {\em purely quadratic} expression with one extra generator:
\begin{theorem}\label{thm2} Let $c_\mu$ for $\mu = 1,\ldots , m$ be as above and let $c_0$ be a  further symbol satisfying
\[ \{ c_0, c_\mu \} = 2\delta_{0\mu}I \hspace{5mm} \mbox{for $\mu = 0,1,\ldots ,m$} \]
extending the set of $c$'s to $m+1$ generators. Introduce
\[ \mbox{$d_0=c_0$ and \hspace{2mm}$d_\mu=ic_\mu c_0 $ for $\mu = 1,\ldots , m$.} \]
(The optional factor $i$ here is just to have all $d$'s hermitian if the $c$'s were.)
Then\\ (a) $ \{ d_\mu,d_\nu \} = 2\delta_{\mu\nu} I\hspace{5mm} \mu , \nu = 0,1, \ldots , m.$\\
(b) A general purely quadratic expression in the $d_\mu$'s for $\mu=0,1, \ldots ,m$ 
\begin{equation}\label{eq1} \tilde{A}= \sum_{\mu ,\nu =0}^m \tilde{a}_{\mu \nu} d_\mu d_\nu \in \cl_2(d{\rm 's}) \end{equation}
is the same as a general quadratic {\em plus linear} expression in the $c_\mu$'s for $\mu=1,\ldots ,m$
\begin{equation}\label{eq2} A=\sum_{\mu ,\nu=1}^m  a_{\mu \nu}\, c_\mu c_\nu +\sum_{\sigma =1}^m b_\sigma c_\sigma\,\,\, \in \cl_{1\oplus 2}(c{\rm 's}).
\end{equation} 
In fact $a_{\mu \nu}=\tilde{a}_{\mu \nu}$ for $\mu, \nu = 1,\ldots ,m$ and $b_\sigma = i(\tilde{a}_{\sigma 0}-\tilde{a}_{0\sigma})$.
\end{theorem}
\noindent {\bf Proof}\, (a) follows immediately from the anti-commutation relations of the $c$'s and the definition of the $d$'s in terms of the $c$'s. For (b) we note that if $\mu ,\nu \neq 0$ then $d_\mu d_\nu = -c_\mu c_0 c_\nu c_0 = c_\mu c_0 c_0 c_\nu = c_\mu c_\nu$ and $d_\mu d_0=ic_\mu c_0 c_0 = ic_\mu$. Inserting these into eq. (\ref{eq1}) gives eq. (\ref{eq2}) with the claimed relations between the coefficients.\, $\Box$

Theorem \ref{thm2} with Remark 1 immediately gives:
\begin{corollary}\label{corol1} $\cl_{1\oplus 2}$ is closed under conjugation by $e^A$ for any $A\in \cl_{1\oplus 2}$.
\end{corollary}

In Section \ref{lie} below we will see an alternative demonstration of this fact, using properties of Lie algebras.

\subsection{ Review of classical simulation of fermionic matchgate circuits}\label{fsim}
Now set $m=2n$ and let $c_\mu$ for $\mu = 1,\ldots ,2n$ be the JW hermitian operators on $n$ qubits. 

\begin{theorem}\label{thm3} Consider any purely quadratic hermitian expression (hamiltonian)
\[ H=i\sum_{\mu ,\nu=1}^{2n} h_{\mu \nu}c_\mu c_\nu \hspace{3mm} \mbox{with $h_{\mu \nu}$ real and anti-symmetric.} \]
Then $U=e^{iH}$ is unitary and\\ (a) all fermionic matchgates $G(V,W)$ (acting on any pair of n.n. lines) with ${\rm det}\, V={\rm det}\, W=1$ arise in this way. For qubit lines $k,k+1$ we use only the corresponding four $c_\mu$'s, having $\mu=2k-1,2k,2k+1,2k+2$.\\ (b) any such $U=e^{iH}$ is expressible as a circuit of 2-qubit fermionic matchgates, of circuit size $O(n^3)$.
\end{theorem}
The proof of this Theorem is given in section 5 of \cite{jm08}. Note that in Theorem \ref{thm3}(a) the case of fermionic matchgates eq. (\ref{gab}) with ${\rm det}\,V={\rm det}\,W \neq 1$ can be readily included by allowing $H$ to also contain quadratic terms $c_\mu c_\nu$ with $\mu = \nu$, thus allowing the identity $II$ on the two qubit lines as a further term in $H$.

Now let $\cc$ be any (poly-sized) circuit of 2-qubit fermionic matchgates, with input a {\em product} state $\ket{\psi_0}$ and output being a final $Z$-measurement on a line $k$, and with no intermediate measurements allowed. If $p_0,p_1$ are the output probabilities then
\[ p_0-p_1= \langle Z_k \rangle = \bra{\psi_0} \cc^\dagger Z_k \cc \ket{\psi_0} \]
where $Z_k$ is $Z$ on the $k^{\rm th}$ line and its expectation value $\langle Z_k \rangle$ is taken in the final state $\cc \ket{\psi_0}$. Now $Z_k=-i c_{2k-1}c_{2k}$ and by Theorem \ref{thm1} (or rather Remark 1) the linear span of pure quadratic terms in the $c$'s is preserved under conjugation by $\cc$. Thus successively conjugating by the 2-qubit gates of $\cc$ (taken in reverse order) we finally arrive at 
\begin{equation} \label{conj} \cc^\dagger Z_k \cc = \sum_{\mu ,\nu=1}^{2n} \alpha_{\mu \nu } c_\mu c_\nu \end{equation}
where the coefficients $\alpha_{\mu \nu }$ can be computed in poly$(n)$ time via successive $2n\times 2n$ matrix multiplications, effecting the conjugation action of the sequence of 2-qubit gates. Then noting that the sum in eq. (\ref{conj}) has only $O(n^2)$ terms and that the $c_\mu c_\nu$'s are product operators, and $\ket{\psi_0}$ is a product state, we see that we can compute $p_0-p_1$ in poly$(n)$ time, giving the efficient classical simulation.

\subsection{Inclusion of 1-qubit gates on the first line (and more)}

Consider now allowing also {\em linear} terms in the hamiltonian
\[ \tilde{H}=i\sum_{\mu ,\nu=1}^{2n} h_{\mu \nu}c_\mu c_\nu + \sum_{\sigma =1}^{2n} k_\sigma c_\sigma \] 
with $k_\sigma$ also real to keep $H$ hermitian. Note that the $e^{i\tilde{H}}$'s include all previous 2-qubit fermionic matchgates as well as further gates which in fact include arbitrary 1-qubit gates $U_1$ on the first qubit line. To see this, recall that the Pauli operators for line 1 are given by \[ X_1=c_1 \hspace{5mm} Y_1=c_2\hspace{5mm} Z_1=-ic_1 c_2 \] so any $U_1= e^{i(\alpha X_1+\beta Y_1 + \gamma Z_1)}$ is now included. There are still further gates involving linear terms in the $c_\sigma$'s with $\sigma > 2$ which generally act across all the first $k$ lines when $\sigma = 2k-1, 2k$ are used.

One way to perform the efficient simulation of these more general circuits is to use the construction in  Theorem \ref{thm2} to reduce the problem to the case of a purely quadratic hamiltonian and then carry out the classical simulation exactly as in Section \ref{fsim}. We can explicitly construct the extra $c_0$ operator by introducing an extra fermionic mode (qubit line) labelled $n+1$ to the right of the existing lines viz. $ 1,2 \ldots , n,n+1$, and set $c_0 = Z \ldots Z X$ where $X$ acts on line $n+1$ and there are $n$ $Z$'s i.e. $c_0$ is just the first JW operator for the new fermionic mode. The previous $2n$  JW operators are all extended by the identity on the new line to recognise the new mode. Thus it is immediate that this $c_0$  satisfies the required anti-commutation relations with $c_\mu$, $\mu =1, \ldots ,2n$. Alternatively we can obtain the extra generator working just within $n$ qubit lines by setting $c_0= Z\ldots Z = (-i)^n c_1c_2 \ldots c_{2n-1}c_{2n}$, which is easily checked to have the required anti-commutation relations.
With either choice of $c_0$ we then
construct $d_\mu$ for $\mu = 0,1, \ldots , n$ as in Theorem \ref{thm2} and as they are still all product operators, we can apply the method of Section \ref{fsim} to achieve the efficient simulation.

The efficient simulation of our more general circuits may also be seen even without introducing the extra operator $c_0$, by using Corollary \ref{corol1} directly -- we just apply the method of Section \ref{fsim} to $\cl_{1\oplus 2}$ replacing $\cl_2$, noting that $\cl_{1\oplus 2}$, like $\cl_2$, has a basis of product operators (viz. the JW $c_\mu$'s and $c_\mu c_\nu$'s) and it is also of polynomial dimension $O(n^2)$.

\noindent {\bf Remark 2.}  Note that this classical simulation method does not depend on $\tilde{H}$ being hermitian and $e^{i\tilde{H}}$ being unitary. Indeed we can replace $\tilde{H}$ by any general complex linear combination $A$ as in eq. (\ref{eq2}) and efficiently compute the quantity $p_0-p_1$ for the corresponding, now non-unitary, circuit of gates $e^A$.  In this setting, general purely quadratic exponents for n.n. lines (cf. Theorem \ref{thm3}(a))  will give gates of the form $G(V,W)$ as in eq. (\ref{gab}) with $V$ and $W$ now being arbitrary invertible $2\times 2$ matrices satisfying ${\rm det}\, V={\rm det}\, W$. \, $\Box$

The fact that $\cl_{1\oplus 2}$ is preserved under conjugation by 1-qubit gates acting on the first line (which provides the key extension beyond the purely non-interacting fermion case) may also be seen by elementary means: any 1-qubit gate can be written as a sequence of products of phase gates $P_\alpha = {\rm diag}(1 \,\,\, e^{i\alpha} )$ and Hadamard gates $H$. Now $P_\alpha \otimes I$ is easily verified to be a fermionic 2-qubit matchgate so we need only consider $H$ on line 1. But $H$ has a very simple conjugation action on the Pauli operators $X,Y$ and $Z$ so its conjugation action on the JW $c_\mu$'s and $c_\mu c_\nu$'s (which are all Pauli products) is very easily computed directly to confirm preservation of $\cl_{1\oplus 2}$.

\subsection{JW formalism for Valiant's simulation theorem}

In this section we show that the simulation above for circuits of exponentials of elements from $\cl_{1\oplus 2}$, includes all of the invertible matchgate cases given in the Main Theorem of \cite{val2} (ibid.  page 1245) i.e. circuits comprising the following kinds of gates:\\
(a) any diagonal 2-qubit matchgate (acting on any pair of qubit lines);\\
(b) any matchgate $B$ acting on n.n. lines, with $B$ having non-zero entries only in the positions $B_{11}$, $B_{22}$,
$B_{33}$, $B_{44}$, $B_{14}$, $B_{41}$, $B_{23}$ and $B_{32}$ (i.e. as in the structure of our fermionic matchgates);\\
(c) any 2-qubit matchgate acting on the first two lines.

According to Theorem \ref{thm11}, any invertible 4x4 matchgate is the exponential of a linear combination of $c_\mu$'s and $c_\mu c_\nu$'s, with $\mu , \nu = 1,2,3,4$ corresponding to the JW operators for two qubit lines. Viewing these two lines as the first two of $n$ lines, we immediately have (c).

For (a) we note that the matchgate identities imply that $B$ is a diagonal matchgate iff $B_{11}B_{44}=B_{22}B_{33}$ so $B$ is the exponential of a linear combination of the commuting matrices $ZI$, $IZ$ and $II$. If the gate acts on lines $k$ and $l$ (with $k<l$) then, noting that $Z_kI_l=c_{2k-1}c_{2k}$, $I_kZ_l=c_{2l-1}c_{2l}$ and $I_kI_l = c_{2k}c_{2k}$ (where $Z_k$ denotes $Z$ acting on the $k^{\rm th}$  line and identity on all other lines etc.), we see that $B$ is the exponential of an element of $\cl_{1\oplus 2}$.

Finally for (b), the matchgate identities imply that any matchgate $B$ satisfying the given non-zero entry conditions, has the form $G(V,W)$ with ${\rm det}\,V={\rm det}\,W$. Then Theorem \ref{thm3}(a) (or more generally its non-unitary extension given in Remark 2) implies that all such gates are again exponentials of elements of $\cl_{1\oplus 2}$, which completes all the cases.

\section{A Lie algebra perspective}\label{lie}
The existence of all of the above efficient classical simulations can also be seen, perhaps even more simply, from some basic Lie algebra theory, as an application of the formalism introduced in \cite{somma}. 
Here we will give an elementary exposition of this view and its application to matchgate circuits.

To motivate this approach, consider first the different and well-studied issue of the efficient classical simulation of Clifford circuits \cite{gottesman, NC, jvdn}. The basic Clifford gates (not to be confused with the Clifford algebras above) are defined to be the Hadamard gate $H$, phase gate $S={\rm diag}\,  (1\,\,  i)$ and the controlled-$Z$ gate $CZ$. A Clifford circuit is any circuit of these gates and a Clifford operation is any such resulting unitary operation on $n$ qubits. Suppose we have a (poly-sized) Clifford circuit on $n$ qubits with overall Clifford operation $C$, and input state $\ket{\psi_{\rm in}} = \ket{\alpha_1} \ldots \ket{\alpha_n}$ being any product state, and output being the result of a standard measurement on the first qubit of the final state $\ket{\psi_{\rm out}}=C\ket{\psi_{\rm in}}$. Then (a variant of) the Gottesman-Knill theorem asserts that this quantum process may be classically efficiently simulated, in the sense that the output probabilities $p_0$ and $p_1$ may be classically computed in poly$(n)$ time.

The key property upon which this result rests, is the fact that Clifford operations conjugate the set of tensor products of 1-qubit Pauli operations into itself i.e. if $P_1,\ldots ,P_n$ are any 1-qubit Pauli operations and $C$ is any Clifford operation then there exist 1-qubit Pauli operations $P_i'$ such that $C^\dagger (P_1\otimes \ldots \otimes  P_n) C = k (P'_1\otimes \ldots \otimes  P'_n)$ where $k=\pm 1$ or $\pm i$. Furthermore the update rule for determining all $n$ of the $P_i'$'s (and $k$) from the $P_i$'s is computable classically in $O(n)$ time if $C$ is a basic Clifford gate. This property then easily gives the classical simulation result for Clifford circuits \cite{cjl, jvdn} viz.
we have
\begin{equation}\label{cliffhanger} p_0-p_1=\langle Z\otimes I\otimes \ldots \otimes I\rangle_{\rm out} =
\bra{\psi_{\rm in}} C^\dagger (Z\otimes I\otimes \ldots \otimes I) C\ket{\psi_{\rm in}}. \end{equation}
Now if the size of the circuit is $N={\rm poly}(n)$ then $C=C_N\ldots C_2C_1$ where each $C_i$ is a basic Clifford gate. So successive conjugation by the $C_i$'s (taken in reverse order) in eq. (\ref{cliffhanger}) gives $P_i'$'s with
\[ p_0-p_1= \bra{\psi_{\rm in}} P_1'\otimes \ldots \otimes P'_n \ket{\psi_{\rm in}} 
=\prod_{i=1}^n \bra{\alpha_i}P_i'\ket{\alpha_i}. \]
Each of the $n$ terms in the latter product can be computed in constant time (ignoring issues of precision which will add at most a poly overhead) and the identities of the $P_i'$'s in $NO(n)$ time so $p_0$ and $p_1$ can be computed in classical $NO(n)+O(n)$ time i.e. poly$(n)$ time.

Now let us isolate the key ingredients that make the above simulation efficient, with a view to generalisation. Consider the following features:\\
{\bf (S1)}: for each $n$ we have a structure $\cs_n$ (above, the Pauli group on $n$ qubits) whose elements have classical poly$(n)$ sized descriptions;\\
{\bf (S2)}: we have a class $\cu_n$ of gates (above, the Clifford gates) that preserve $\cs_n$ under conjugation, and the conjugation update rule is computable in classical poly$(n)$ time;\\
{\bf (S3)}: for a suitable class of input states $\ket{\psi_{\rm in}}$ (say product states or computational basis states) we have $\bra{\psi_{\rm in}} A \ket{\psi_{\rm in}}$ being computable in poly$(n)$ time for any $A\in \cs_n$;\\
{\bf (S4)}: $\cs_n$ contains observables of interest e.g. $Z\otimes I \otimes \ldots \otimes I$.

Clearly if these features are satisfied then we will have a classical simulation result for circuits of gates from $\cu_n$ and expectation values of observables $A\in \cs_n$ (that we have used above to obtain our output probabilities). Note that $\cs_n$ need not be a group; in fact in \cite{jozbeth} it has been pointed out that the JW simulation of fermionic matchgate circuits can be viewed as an example of the above features with $\cs_n$ being a vector space (of linear (or quadratic) terms in a Clifford algebra). But more generally we seek further natural occurrences of these features in other mathematical contexts, with an aim of identifying new classes of classically simulatable quantum circuits.

In our motivating discussion above (and throughout the paper) we are considering only non-adaptive circuits i.e. we do not allow intermediate measurements within the circuit, followed by adaptive choices of subsequent gates depending on earlier measurement outcomes. Furthermore, for matchgate circuits we consider only unitary circuits, without intermediate measurements, having measurements only at the end to provide output probabilities. 
The computational power of such unitary matchgate circuits has been 
shown in~\cite{jkmw} to coincide with that of space-bounded quantum 
computation, and further results on the ability of these circuits to 
compute Boolean functions have been given in~\cite{vdn2011}.
Some classical simulation results for adaptive matchgate circuits have been given in~\cite{terdiv}. In~\cite{jvdn} the simulation complexity for adaptive and non-adaptive Clifford circuits has been discussed. It is shown there that non-adaptive Clifford circuits with product state inputs can be classically efficiently simulated (essentially by the method above) whereas if the circuits are allowed become adaptive then they become universal for quantum computation. (The latter result depends on special properties of Clifford gates, such as the fact that CNOT is Clifford.) This suggests that the formalism we are developing here may not have any generic extension to the case of adaptive circuits.

Somma et al. \cite{somma} have identified an occurrence of the features (S1) - (S4) above arising naturally in the theory of representations of Lie algebras and Lie groups. Here we will give an exposition of this formalism and describe how it relates to our original issue of the classical simulation of matchgate circuits. We will not need abstract Lie algebra theory and we begin with the concrete setting of a {\em finite-dimensional matrix Lie algebra} $\ca$ viz. a vector space $\ca$ of matrices (of some finite dimension $d$, not to be confused with the size of the matrices) that is closed under the commutator (or bracket operation) $[A,B]=AB-BA$ (defined in terms of the usual matrix product). If $B_1,\ldots ,B_d$ is a basis of matrices for $\ca$ then we have the associated {\em structure constants} $c^k_{ij}$ given by
\[ [B_i,B_j]=\sum_{k=1}^d c^k_{ij}B_k. \] 

%
Introduce the set of all exponentials ${\cal E} = \{e^A: A\in {\cal A}\}$ (where the matrix exponential is the sum $\sum_{k=0}^\infty A^k/k$). Note that all elements of $\cal E$ are invertible and $\cal E$ is closed under inverses. Let $\cal G$ be the matrix group generated by $\cal E$. $\cal G$ is in fact a Lie group with Lie algebra $\cal A$ (cf.~\cite{fh}~\textsection{8.3}) but we will not explicitly need this fact here -- for our key result, Lemma~\ref{adjoint} below, it will suffice to know just that $\cal A$ is closed under commutators. However it is interesting to note that in a more abstract setting of Lie group representation theory, Lemma~\ref{adjoint} amounts to an instance of a fundamental general result (cf.~\cite{fh}) viz. that the Lie algebra $\cal A$ carries a natural representation of the Lie group $\cg$. This is the adjoint representation in which each $G\in \cg$ acts linearly on $\ca$ by conjugation, and for us an important ingredient of this result is the fact that $\ca$ is always preserved under conjugation by such $G$'s:

\begin{theorem} (Adjoint representation of a Lie group on its Lie algebra.) With $\ca$ and $\cg$ as above, for all $G\in \cg$ and $B\in \ca$ we have that $B'=G BG^{-1}$ is in $\ca$ and the resulting linear map $B\rightarrow B'$ on $\ca$ for each $G\in \cg$, provides a representation of $\cg$ on $\ca$.
\end{theorem}
We will further need to assess the complexity of computing the update $B\rightarrow B'$, as given in the following Lemma (whose proof also implicitly shows that $\cal A$ is preserved under conjugation by $G\in\cg$).

\begin{lemma}\label{adjoint} The conjugation action of $e^A$ on $\ca$ can be classically computed (to $m$ digits of accuracy) in poly$(m,d)$ time (where $d$ is the dimension of $\ca$).
\end{lemma}

\noindent {\bf Proof}\, Let $B_1,\ldots ,B_d$ be a basis for $\ca$ and write $A=\sum \xi_j B_j$. We aim to compute $a_{ij}$ defined by $e^AB_ie^{-A}=\sum_{ij}a_{ij}B_j$ which will suffice to fully characterise the adjoint action of $e^A$. To this end, introduce 
\[ B_i(t)=e^{tA}B_ie^{-tA} \] so $B_i(0)=B_i$ and we get
\[ \frac{dB_i(t)}{dt}=[A,B_i(t)] =\sum_{jk}\xi_jc_{ji}^kB_k(t) \]
so if $\underline{B}(t)=(B_1(t),\ldots ,B_d(t))^T$ then
\[ \frac{d\underline{B}(t)}{dt}=M\underline{B}(t)\hspace{5mm} \mbox{with}\hspace{5mm} 
M^k_i=\sum_j \xi_jc_{ji}^k \]
and so $\underline{B}(t)=e^{Mt}\underline{B}(0)$. Finally setting $t=1$ we obtain the matrix of values $[a_{ij}]$ as $[a_{ij}]=e^M=I+M+M^2/2!+\ldots$. The exponential series converges rapidly and the result  (involving $d\times d$ matrix algebra) can be computed to $m$ digits with $O(m)$ terms, so the whole calculation takes poly$(m,d)$ time.\, $\Box$

Now with the above in view, we can make a connection to our desired features (S1) -- (S4). For each $n$, to $n$ qubit lines we associate a matrix Lie algebra $\ca_n$ of dimension $d=$poly$(n)$, comprising matrices of size $2^n\times 2^n$. If furthermore, $\ca_n$ has a basis of matrices that are tensor products of 1-qubit matrices (hence having  poly$(n)$ sized descriptions) then (S1) will be satisfied. If $A\in \ca_n$ is skew-hermitian then $e^A$ will be unitary and we take $\cu_n$ to be the corresponding set of unitary operations. Lemma \ref{adjoint} then guarantees that (S2) will be satisfied. (Note that Lemma \ref{adjoint} applies for all $A\in\ca$, even if $e^A$ is not unitary, and this leads to a classical simulation result for circuits of gates that are not necessarily unitary.) For (S3) and (S4) we just choose suitably well behaved classes; for example (as will apply in our case below), if $\ca_n$ has a basis of product matrices we can take input states to be product states, and require also that $\ca_n$ contains say $I\otimes \ldots\otimes I\otimes Z\otimes I \ldots \otimes I$ (having $Z$ on the $k^{\rm th}$ line), for some or all $1\leq k\leq n$.

Finally let us return to the JW simulation of matchgate circuits, to see it as an example of the above formalism. Consider again $2n$ generators $c_\mu$, $\mu=1,\ldots ,2n$ that satisfy the Clifford algebra anti-commutation relations eq. (\ref{cliffcomm}) and  we may concretely regard them as being the Jordan-Wigner operators eq. (\ref{jwr}). The linear span $\cl_1$ of the $c_\mu$'s is not closed under commutators e.g. if $\mu\neq \nu$ then $[c_\mu,c_\nu]=c_\mu c_\nu -c_\nu c_\mu=2c_\mu c_\nu$ which is not generally expressible as a linear sum. However the linear span $\cl_2$ of all quadratic products $c_\mu c_\nu$ {\em does} form a Lie algebra, being closed under commutators by virtue of the Clifford {\em anti}-commutation relations; indeed we have $[c_\mu c_\nu\,   ,\, c_\alpha c_\beta  ] = c_\mu c_\nu c_\alpha c_\beta - c_\alpha c_\beta c_\mu c_\nu$ which is zero if all indices are distinct, or it reduces to a quadratic expression again if two indices are equal. If we omit the identity matrix from $\cl_2$ (i.e. the case of $\mu = \nu$ in quadratic terms) then the remaining
 Lie algebra has dimension $d=  \binom{2n}{2}=n(2n-1)=O(n^2)$. This algebra is in fact isomorphic to the (complexified) Lie algebra of the special orthogonal group SO$(2n)$ in $2n$ dimensions, as for example, they have the same structure constants for suitable choices of bases. (The quadratic terms with $\mu =\nu$ then just contribute an extra single dimension to the Lie group as an overall scalar multiple for the matrices $e^A$). Thus taking $\ca_n$ for $n$ qubit lines to be the Lie algebra $\cl_2$ (and using the JW operators for the $c_\mu$'s) we have (S1) -- (S4) all holding and we obtain our efficient classical simulation of fermionic matchgate circuits viz. circuits of gates that are exponentials of quadratic expressions in the JW operators.

In the same way, from the Lie algebra formalism we can also easily obtain the classical simulation result for hamiltonians that involve both quadratic and linear terms i.e. gates $e^A$ with exponents $A\in \cl_{1\oplus 2}$. For this we simply notice that (despite that fact that $\cl_1$ is not a Lie algebra), $\cl_{1\oplus 2}=\cl_1\oplus \cl_2$ is a Lie algebra, again closed under commutators by virtue of the Clifford algebra anti-commutation relations. If we again omit the case of $\mu =\nu$  in the quadratic terms then the resulting Lie algebra has dimension $d=\binom{2n}{2}+2n =n(2n+1)=O(n^2)$, and it is isomorphic to the (complexified) Lie algebra of the orthogonal group SO$(2n+1)$.

It would be interesting to seek further natural occurrences of the conditions in (S1) -- (S4), perhaps using other Lie algebras, or indeed further unrelated constructions, recalling that our original motivating example of Clifford circuits does not itself seem to arise from any underlying Lie algebra (as Clifford operations form only a discrete set of gates).

\section{Appendix: proof of Theorem \ref{thm11}}

Before proving this theorem we introduce some further terminology. Note first that for each $ij$, $B_{ij}$ occurs in exactly five of the matchgate identities. Let $\cm (ij)$ be the corresponding set of five identities and let $\cn (ij)$ be the remaining five. It was noted in \cite{ccl} that the set of matchgate identities possesses a high degree of symmetry. This leads to associated dependencies and in fact, for any non-zero matrix, only five of the ten matchgate identities are significant, in the following sense.
\begin{lemma} \label{lemma1} Let $ij$ be given. Suppose that a matrix $B$ has $B_{ij}\neq 0$ and $B$ satisfies the identities in $\cm(ij)$. Then $B$ satisfies the identities in $\cn(ij)$ too, so $B\in \cm\cg$.
\end{lemma}
\noindent {\bf Proof.}\, Consider the illustrative case of $ij=44$, which has $\cm(44)=\{ M_1,M_2,M_3,M_4,M_5 \}$ and $\cn(44)=\{ 
M_6,M_7,M_8,M_9,M_{10} \}$. Now consider each element of $\cn(44)$ in turn, multiplied by $B_{44}$. For $M_6$ we have:
\[ B_{44}M_6= (B_{11}B_{44})B_{24}-(B_{12}B_{44})B_{23}+(B_{13}B_{44})B_{22}-(B_{21}B_{44})B_{14}. \]
Each bracketed term is the $B_{44}$-term of an identity $M_k$ in the set $\cm(44)$ viz. respectively $M_1,M_5,M_4$ and $M_2$. Replacing the brackets by these full expressions we find that all the extra terms cancel and we get:
\[  B_{44}M_6= (M_1)B_{24}-(M_5)B_{23}+(M_4)B_{22}-(M_2)B_{14}. \]
So if $B_{44}\neq 0$ then $M_1= \ldots =M_5=0$ implies that $M_6=0$. The same procedure works for all other $M_i$'s in $\cn(44)$ too. Furthermore it also works for any $\cm(ij)$ and $\cn(ij)$ (for all initial choices of $ij$). The mind-numbingly long list of (eighty) claimed algebraic relations can be readily verified, for example by computer algebra.\ $\Box$

\noindent {\bf Proof of Theorem \ref{thm11}(a).}\, We need to show that any non-invertible matchgate $B$ is the limit of invertible matchgates. If $B$ is the all-zero matrix, let $\tilde B$ be any invertible matchgate and setting $\tilde B_k=\tilde B/k$ we have $B=\lim_{k\rightarrow \infty} \tilde B_k$.
Thus suppose that $B$ contains some non-zero entry $B_{i_0j_0}=c\neq 0$, which will remain constant in the following constructions. Let $B_{i_1j_1},\ldots ,B_{i_5j_5}$ be the multipliers of $B_{i_0j_0}$ in the five matchgate identities of $\cm(i_0j_0)$. Dividing these identities by $c$ we obtain 
$B_{i_1j_1},\ldots ,B_{i_5j_5}$ expressed in terms of $c$ and the ten entries $B_{kl}$ with $kl \neq i_0j_0, i_1j_1,\ldots ,i_5j_5$. 

We substitute these into $B$ and for fixed $B_{i_0j_0}=c$ we obtain a matrix $\tilde B$ that is freely parameterised by ten complex variables (viz. the $B_{kl}$ above) and which satisfies $\cm(i_0j_0)$ with $B_{i_0j_0}=c \neq 0$. Thus by Lemma \ref{lemma1}, $\tilde B \in \cm\cg$. 

Then $\det \tilde B$  is a polynomial in the ten variables. Now for any $i_0j_0$ there is an invertible matchgate whose $i_0j_0^{\rm th}$ entry is $c$ so $\det \tilde B$ cannot be identically zero. Thus its zero set must have empty interior so $\{\tilde B: \det \tilde B \neq 0 \}$ is dense in the set of all matchgates $\tilde B$ with $\tilde B_{i_0j_0}=c$. Hence $B$ is the limit of such invertible matchgates.\, $\Box$

To facilitate the proof of Theorem \ref{thm11}(b) we first establish a geometrical interpretation of the ten matchgate identities, which will also provide a connection to the Jordan-Wigner operators. It is an extension of a construction in \cite{knill}, given there for a setting of five matchgate identities.

We begin by introducing the four-qubit vector
\[ \ket{\Upsilon}= \sum_{j=1}^4 \ket{j}\ket{j} \]
(where the labels $j= 1,2,3,4$ correspond to qubit labels $00,01,10,11$ respectively). Then  (up to an overall factor of $i$) we introduce the so-called Choi-Jamiolkowski state for the 2-qubit operation $XY$ (writing $I$ for the identity on the first two qubits):
\[ \ket{F_0} = I\otimes (XY) \ket{\Upsilon} = \ket{1}\ket{4}-\ket{2}\ket{3}+\ket{3}\ket{2}-\ket{4}\ket{1} \]
which is an {\em anti-symmetric} vector in $\CC^4 \otimes \CC^4$. We extend $\ket{F_0}$ to a full orthogonal basis of the 6-dimensional anti-symmetric subspace with the following vectors:
\begin{align*}
&|F_0\ra=|1\ra|4\ra-|4\ra|1\ra-|2\ra|3\ra+|3\ra|2\ra \\
&|F_1\ra =|1\ra|4\ra-|4\ra|1\ra+|2\ra|3\ra-|3\ra|2\ra \\
&|F_2\ra = |1\ra|2\ra-|2\ra|1\ra\\
&|F_3\ra=|1\ra|3\ra-|3\ra|1\ra\\
&|F_4\ra=|2\ra|4\ra-|4\ra|2\ra\\
&|F_5\ra=|3\ra|4\ra-|4\ra|3\ra.
\end{align*}
Since all these vectors are real we have $\bra{F_i}=\ket{F_i}^\dagger= \ket{F_i}^T$ (where $T£$ denotes transpose and $\dagger$ the conjugate transpose). Now for any $4\times 4$ matrix $B$, $B\otimes B$ preserves the anti-symmetric subspace. Introduce
\[ \begin{array}{rcl}
D_i &=& \bra{F_i} (B\otimes B) \ket{F_0}\\
D_i^T &=& \bra{F_0} (B\otimes B) \ket{F_i} = \bra{F_i}B^T\otimes B^T \ket{F_0}
\end{array} \]
By direct calculation it is easy to verify the following relations:
\begin{align*}
D_1+D_1^T &= 4M_1\\
D_4 &= 2M_2\\
D_5 &= 2 M_3\\
D_5^T &= 2M_4\\
D_4^T &= 2M_5\\
D_2 &= 2M_6\\
D_2^T &= 2M_7\\
D_1-D_1^T &= 4M_8\\
D_3 &= 2M_9\\
D_3^T &= 2M_{10}
\end{align*}
Thus $B$ is a matchgate if and only if $D_i=0$ and $D^T_i=0$ for $i=1, \ldots ,5$. Now since $\{ \ket{F_i}:i=0,1, \ldots ,5 \}$ is an orthogonal basis for the anti-symmetric subspace we see that $D_i=0$ for $i=1,\ldots ,5$ iff $\ket{F_0}$ is an eigenvector of $B\otimes B$. Similarly 
$D_i^T=0$ for $i=1,\ldots ,5$ iff $\ket{F_0}$ is an eigenvector of $B^T\otimes B^T$ and we have proved:
\begin{theorem}\label{thm22} A $4\times 4$ matrix $B$ is a matchgate iff $\ket{F_0}$ is an eigenvector of both $B\otimes B$ and $B^T\otimes B^T$.
\end{theorem}
We remark that Theorem \ref{thm22} immediately implies that the set of invertible matchgates forms a group, which was proven by other means in \cite{ccl}.

\noindent {\bf Proof of Theorem 1(b).}\, We show that $\cg \subseteq \cm\cg^*$ and $\cm\cg^*\subseteq \cg$. For the first inclusion consider again the 11 generators of the Lie algebra $\cl$ given in eq. (\ref{liealg}), which we now label as $A_0=II$ and $A_i$ with $i=1, \ldots ,10$ for the others. It is easy to check that
\begin{equation}\label{tanti} A_i (XY)+(XY)A^T_i=0\hspace{3mm} \mbox{for $i=1,\ldots ,10$.} \end{equation}
Next we recall that $\ket{F_0}=I\otimes (XY) \ket{\Upsilon}$ and note the following facts: (i) for any 2-qubit operator $W$ we have $I\otimes W \ket{\Upsilon}=W^T\otimes I \ket{\Upsilon}$ and (ii) $I\otimes W \ket{\Upsilon}=0$ iff $W=0$. Using these facts we can see that eq. (\ref{tanti}) is equivalent to 
\begin{equation}\label{fcond} 
(A_i \otimes I+I\otimes A_i) \ket{F_0}=0\hspace{3mm} \mbox{for $i=1,\ldots ,10$.}
\end{equation}
Now if $A=\sum_{i=0}^{10} \alpha_i A_i$ is any element of the Lie algebra $\cl$ we have
\[ e^{(A\otimes I+I\otimes A)}=  e^{A\otimes I}e^{I\otimes A}= B\otimes B\hspace{3mm} \mbox{where $B=e^A$.} \]
Then eq. (\ref{fcond}) (and the fact that $A_0$ is the identity operation)  implies that $\ket{F_0}$ is an eigenvector of $e^{(A\otimes I+I\otimes A)}$ i.e. of $B\otimes B$. Since $\cl$ is closed under taking transposes and $B^T=e^{(A^T)}$, we similarly have $\ket{F_0}$ being an eigenvector of $B^T\otimes B^T$, so by Theorem \ref{thm2}, $B\in \cm\cg^*$ and $\cg\subseteq \cm\cg^*$.

For the reverse inclusion, let $B\in \cm\cg^*$ be any invertible matchgate. Then (since $B$ is invertible) $B=e^A$ for some  $4\times 4$ matrix $A$ and $B\otimes B= e^A\otimes e^A$ so Theorem \ref{thm2} gives
\[ e^A\otimes e^A \ket{F_0}=\lambda \ket{F_0} \]
for some $\lambda \neq 0$. Thus $e^{tA}\otimes e^{tA} \ket{F_0}=\lambda^t \ket{F_0}$ for $t\in \RR$ and taking $\frac{d}{dt}|_{t=0}$ we get
\[ (A\otimes I+I\otimes A) \ket{F_0}=\lambda'\ket{F_0} \]
for some $\lambda'$. So
\begin{equation}\label{Afive}    \bra{F_i}A\otimes I+I\otimes A)\ket{F_0}=0\hspace{3mm} \mbox{for $i=1,\ldots ,5$.}\end{equation}
This gives five linear equations on the sixteen entries of $A$. Since $\ket{F_i}$ are orthogonal, the equations are independent, and we must have an 11-dimensional linear space of solutions. Now any $4\times 4$ matrix $A$ can be written as
\[ A=\sum_{i,j=0}^3 \alpha_{ij}P_i\otimes P_j \]
where $P_0=I$,  $P_1=X$, $P_2=Y$ and $P_3=Z$ are the Pauli matrices. We know from eq. (\ref{fcond}) (and $I\otimes I \ket{F_0}=\ket{F_0}$) that 
all 11 generators $A_0,A_1,\ldots ,A_{10}$ of the Lie algebra $\cl$ satisfy eq. (\ref{Afive}) so $\cl$ itself must be the 11-dimensional linear space of solutions
of eq. (\ref{Afive}) i.e. $A\in \cl$ so $B=e^A \in \cg$, completing the proof of Theorem \ref{thm11}(b).\, $\Box$

Finally we mention a possible alternative ``brute force'' approach to proving the inclusion $\cg \subseteq \cm\cg^*$. Any element of $\cg$ has the form $e^A$ where $A$ is a (complex) linear combination of the eleven $4\times 4$ matrices given explicitly in eq. (\ref{liealg}). Thus we could envisage using computer algebra to explicitly compute $e^A$ symbolically as a function of eleven variables and then check each of the matchgate identities on the resulting matrix elements. A significant issue here is the complexity of the symbolic manipulations needed to simplify the very long multivariate algebraic expressions obtained. Using straightforward programming in Mathematica implemented on a standard modern laptop, it took eight hours to compute and simplify all sixteen entries of $e^A$ and many of the matchgate identities required several more hours each, for their explicit symbolic verification on the resulting matrix entry expressions. We would expect that these timings could probably be significantly reduced by more perspicacious programming.

\subsection*{Acknowledgments}
Thanks to Niel de Beaudrap and Dan Browne for helpful discussions and to Leslie Valiant for raising the question of the full equivalence of matchgates and the JW formalism. RJ was supported in part by the EC network Q-ALGO. AM was supported in part by National Science Foundation grants PHY-1212445 and PHY-1314955.


\begin{thebibliography}{~~} \label{refs}

\bibitem{val2} L. Valiant Quantum circuits that can be simulated classically
in polynomial time. {\em SIAM J. Computing} {\bf 31:4}, 1229-1254 (2002).

\bibitem{val1} L. Valiant,   Holographic algorithms. {\em SIAM J. Computing}
{\bf 37:5} 1565-1594 (2007).

\bibitem{terdiv}B. Terhal and  D. DiVincenzo,  Classical simulation of
noninteracting-fermion quantum circuits. {\em Phys. Rev. A} {\bf
65}, 032325/1-10 (2002).

\bibitem{knill}E.  Knill,  Fermionic linear optics and matchgates. Preprint available at
arXiv:quant-ph/0108033 (2001).

\bibitem{brkit} S. Bravyi and A. Kitaev, Fermionic quantum computation. {\em Annals of Physics} {\bf 298}, Iss. 1  pp.210-226 (2002)

\bibitem{jm08} R. Jozsa and  A. Miyake, Matchgates and classical simulation of quantum circuits.
{\em Proc. R. Soc. (Lond) A} {\bf 464}, p3089-3106 (2008).

\bibitem{somma} R. Somma, H. Barnum, G. Ortiz and E. Knill, 2006 Efficient
solvability of hamiltonians and limits on the power of some
quantum computational models. {\em Phys. Rev. Lett.} {\bf 97},
190501.

\bibitem{jw}P. Jordan and E. Wigner,  \"{U}ber das Paulische
\"{A}quivalenzverbot. {\em Zeitschrift f\"{u}r Physik} {\bf 47},
631-651 (1928).

\bibitem{ccl} J-Y. Cai, V. Choudhary, P. Lu, On the Theory of Matchgate Computations, ccc, pp.305-318, {\em Twenty-Second Annual IEEE Conference on Computational Complexity (CCC'07)}  (2007).

\bibitem{NC} M.  Nielsen and I. Chuang, Quantum
Computation and Quantum Information. Cambridge University Press (2000).

\bibitem{gottesman} D. Gottesman, 
Stabilizer Codes and Quantum Error Correction, PhD thesis,
California Institute of Technology, Pasadena, CA (1997).

\bibitem{jvdn} R. Jozsa and M. Van den Nest, Classical simulation complexity of extended Clifford circuits, 
{\em Quant. Inform.  Comp.}  {\bf 14} p633-648 (2014).

\bibitem{cjl} S. Clark, R. Jozsa and N. Linden,  Generalised Clifford groups and simulation of associated quantum circuits, {\em Quant.
Inform.  Comp.}  {\bf 8} p106-126 (2008)

\bibitem{jozbeth} R. Jozsa, Embedding classical into quantum computation, {\em Springer LNCS} {\bf
5393} Beth Festschrift, J. Calmet, W. Geiselmann, J. Mueller-Quade
(eds.), p43-49 (2008).

\bibitem{fh} W. Fulton and J. Harris, Representation theory: a first course. Graduate Texts in Mathematics 129, Springer-Verlag New York (1991).

\bibitem{jkmw} R. Jozsa, B. Kraus, A. Miyake and J. Watrous, Matchgate 
and space-bounded quantum computations are equivalent. {\em Proc. R. 
Soc. (Lond) A} {\bf 466}, p809-830 (2010).

\bibitem{vdn2011} M. Van den Nest, Quantum matchgate computations and 
linear threshold gates. {\em Proc. R. Soc. (Lond) A} {\bf 467}, p821-840 (2011).






\end{thebibliography}
\end{document}